\documentclass[twocolumn]{aastex62}

\usepackage{amsmath}
\usepackage{color}
\usepackage{graphicx}
\usepackage{xcolor}
\usepackage{hyperref}
\hypersetup{pdfauthor=Judit Szul\'agyi}
\hypersetup{backref=true, pagebackref=true, hyperindex=true, breaklinks=true,colorlinks=true,urlcolor=blue, linkcolor=blue,  citecolor=blue,pagecolor=red, bookmarks=true, bookmarksopen=true}

\received{February 23, 2020}
\revised{XX, 2020}
\accepted{XX, 2020}
\submitjournal{ApJ}

\shorttitle{Hydrogen recombination lines from planets}
\shortauthors{Szul\'agyi \& Ercolano}

\begin{document}

\title{Hydrogen recombination line luminosities and -variability from forming planets}

\correspondingauthor{J. Szul\'agyi}
\email{judit.szulagyi@uzh.ch}

\author[0000-0001-8442-4043]{Judit Szul\'agyi}
\affil{Center for Theoretical Astrophysics and Cosmology, Institute for Computational Science, University of Z\"urich,\\
Winterthurerstrasse 190, CH-8057 Z\"urich, Switzerland.}

\author[0000-0001-7868-2740]{Barbara Ercolano}
\affil{Universit\"ats-Sternwarte M\"unchen, Scheinerstr. 1, D-81679 M\"unchen, Germany \\ 
Excellence Cluster Origin and Structure of the Universe, Boltzmannstr.2, D-85748 Garching bei M\"unchen, Germany.}

\begin{abstract}
We calculated hydrogen recombination line luminosities (H-$\alpha$, Paschen-$\beta$ and Brackett-$\gamma$) from three dimensional thermo-hydrodynamical simulations of forming planets from 1 to 10 Jupiter-masses.  We explored various opacities to estimate the line emissions with extinction in each cases assuming boundary layer accretion. When realistic opacities are considered, only lines from planets $\ge$10 Jupiter-mass can be detected with current instrumentation, highlighting that from most planets one cannot expect detectable emission. This might explain the very low detection rate of H-$\alpha$ from forming planets from observations. While the line emission comes from both the forming planet and its circumplanetary disk, we found that only the disk component could be detected due to extinction. We examined the line variability as well, and found that it is higher for higher mass planets. Furthermore, we determine for the first time, the parametric relationship between the mass of the planet and the luminosity of the hydrogen recombination lines, as well as the equation between the accretion luminosity and hydrogen recombination line luminosities.
\end{abstract}

\keywords{accretion, accretion disks --- methods: numerical --- planets and satellites: formation --- planets and satellites: gaseous planets --- protoplanetary disks}

\section{Introduction}
 \label{sec:intro}

Hydrogen recombination lines are traditionally used in the astrophysics community to estimate accretion rates. Similarly to young stars, forming planets are also believed to emit Hydrogen recombination lines, however it is still unclear, what planetary masses can indeed cause detectable hydrogen ionization. Two processes are possible to cause hydrogen ionization: (i) thermal ionization occurring for gas temperatures above $\sim$10,000 K, or (ii) collisional ionization, requiring an accretion flow fast enough to produce ionized hydrogen. Whether these conditions present for planetary mass objects, that are cooler than stars, is still subject of current investigations \citep{AB09a,AB09b,Szulagyi17}. 
The first detection of an accreting planet via the H-$\alpha$ tracer in the circumstellar disk (CSD) was in the system of LkCa15 \citep{Sallum15}. However, follow-up observations of this source by independent groups were unable to detect the H-$\alpha$ signal again  \citep{Mendigutia18}. More recently, H-$\alpha$ was detected from the planetary candidate PDS 70b, with an estimated accretion rate of $10^{-8} \rm{M_{Jup}/year}$  \citet{Wagner18}. The detection of H-$\alpha$ from PDS 70b was confirmed in 2019 and, additionally, another planet-like H-$\alpha$ source was found in the system \citet{Haffert19}.
It is however puzzling that despite the effort from ongoing surveys \citep{Cugno19,Zurlo20} aiming at detecting forming planets via H-$\alpha$, until now planet-like H-$\alpha$ detections have only been obtained for the above two circumstellar disks. In the \citet{Zurlo20} survey, the authors could close out accretion luminosity of $10^{-6} \rm{L_{\odot}}$ at a separation of $0."2$ from the host star. In  \citet{Cugno19} the upper limit for H-$\alpha$ fluxes was around $10^{-14} -- 10^{-15} erg/s/cm^2$ for the different systems. Given that H-$\alpha$ is generally a robust tracer of accretion, the question then arises as to why are there so few defections of this line from forming planets. One problem may be the extinction. The forming planet is surrounded by a circumplanetary disk \citep{Kley99,Lubow99}, which may absorb most of the hydrogen recombination lines emitted by the accreting planets \citep{Szulagyi19}. At medium to large inclinations, the circumstellar disk can also absorb out a significant fraction of the H-$\alpha$ flux. This complex three-dimensional problem has prevented until now the magnitude of the extinction to be robustly estimated. The second problem is clearly that planets are not as hot as stars, and even the accretion flow to the forming planet cannot always heat up the gas and ionize it. It is important to understand what is the temperature on the surface of the forming planet and in the circumplanetary disk in order to investigate whether and where hydrogen ionization can occur. The third possible problem why there have not been many of H-$\alpha$ detections from forming planets is the potential variability of this line. As planets orbit in the circumstellar disks, their accretion rate and the extinction rate changes as well, presumably causing variations in the hydrogen recombination line luminosities. The magnitude of the possible line variability is currently unknown. 

The first rough estimates of H-$\alpha$ luminosities from 3D thermo-hydrodynamical simulations of forming planets and their circumplanetary disks, used the T Tauri empirical formula \citep{Rigliaco} that connects $L_{acc}$ to $L_{H\alpha}$ \citep{SzM17}. These models ignored extinction and suggested that all examined planetary mass (1, 3, 5, 10 $\rm{M_{Jup}}$) might emit of the order of 4 to $7\times 10^{-6}\,\rm{L_{Sun}}$ in $L_{H\alpha}$ \citep{SzM17}. This result is in strong contrast to the low detection rate of H-$\alpha$ from observational surveys. Part of the discrepancy may arise from the fact that the accretion process around planets could be significantly different from that of stars, rendering the T Tauri empirical formula inadequate in this case. Indeed, while stars have strong magnetic fields which lead to magnetospheric accretion, planets are expected to have much weaker fields (at least today they have approximately three orders of magnitude smaller field-strengths) , so in many cases not strong enough to support magnetospheric accretion. It is suspected that planets may instead grow via boundary layer accretion, when the circumplanetary disk directly touches the planet surface \citep{OM16}, and this is why we used this assumption. However, there is no clear concensus yet which way forming planet's accrete, due to lack of adequate magneto-hydrodynamic simulations of forming planets.

\begin{figure*}
\includegraphics[width=\textwidth]{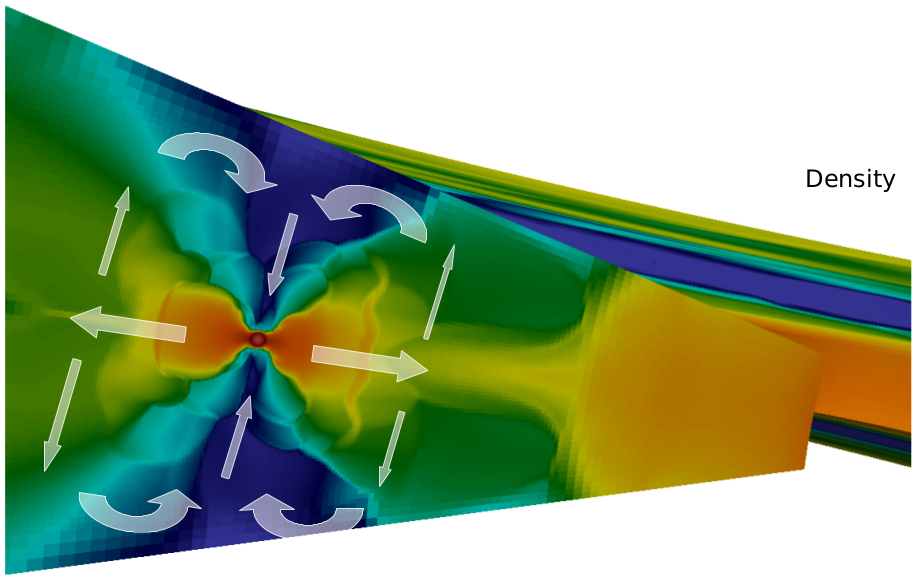}
\caption{Meridional circulation (from \citealt{Szulagyi14} between the circumstellar and circumplanetary disk (orange "butterfly-pattern" in the center of the image) schematized on a 2D vertical surface for simplicity. Gas from the higher density regions enters the lower density gap regions, then spirals down to the CPD. The non-accreted gas is then pushed back inside the CPD midplane regions to the circumstellar disk. It rises up again there to maintain the vertical hydrostatic equilibrium and to maintain the flow. This continuous material feeding from the circumstellar disk assures that the CPD density scales with the CSD density \citep{Szulagyi17}. Streamlines video are available at: \url{https://www.ics.uzh.ch/~judits/images/visu/video_6.mp4}}
\label{fig:merid}
\end{figure*}

\begin{figure*}
\includegraphics[width=\textwidth]{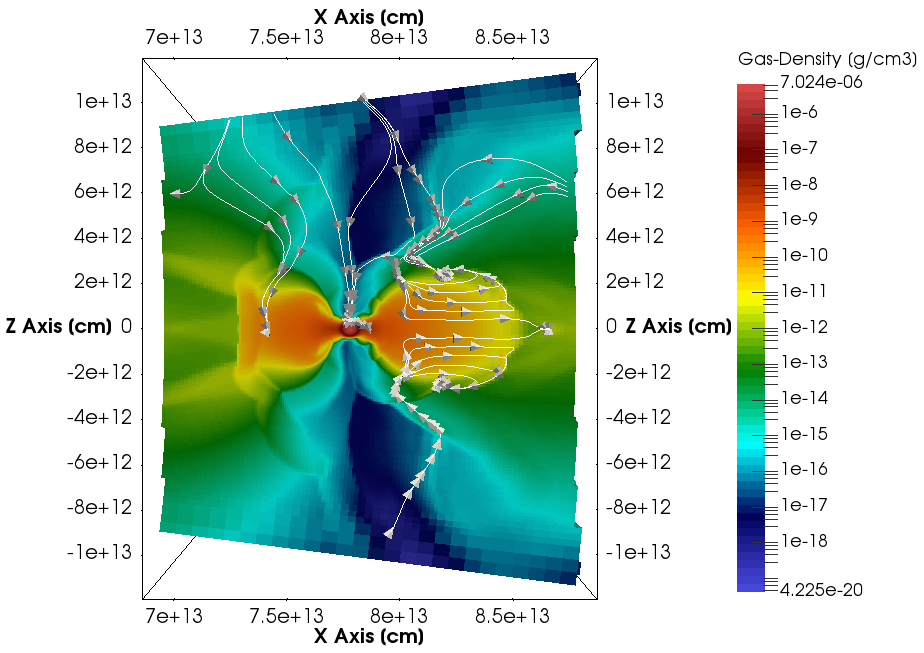}
\caption{Zoom into the circumplanetary disk in the 10 $\rm{M_{Jup}}$ simulation, with a few representative streamlines of the gas to show the flow direction. Some of the streamlines accrete directly onto the planet, while others, further away from the planet in the x-direction, land onto the circumplanetary disk, and leave it in the midplane regions to flow back to the circumstellar disk thus maintaining meridional circulation \citep{Szulagyi14,FC16}.}
\label{fig:streams}
\end{figure*}

\begin{figure*}
\includegraphics[width=\textwidth]{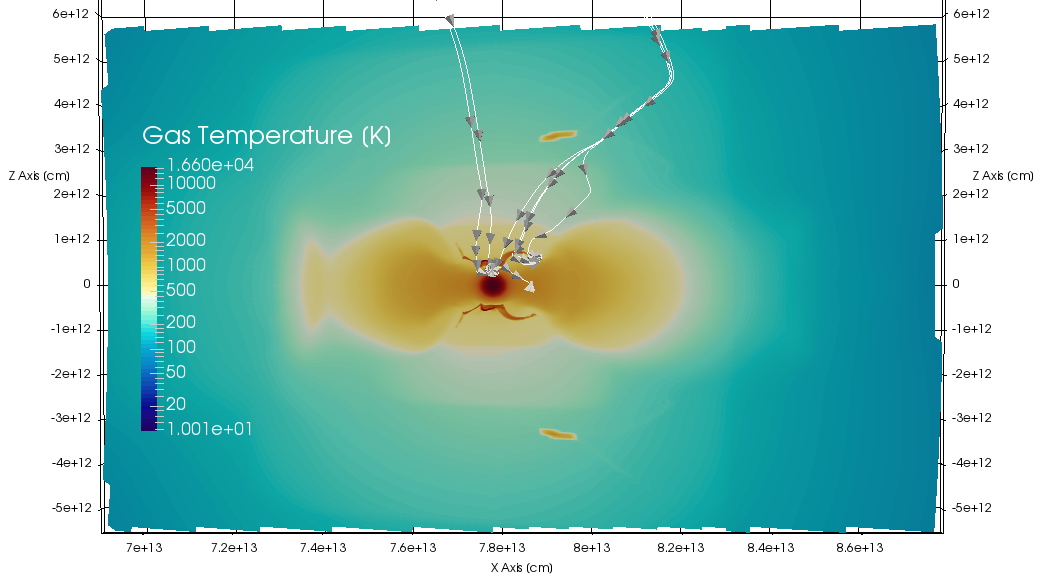}
\caption{Temperature-map of the circumplanetary disk in the 10 $\rm{M_{Jup}}$ simulation (i.e. the hottest simulation). As mass flows in with high velocities (few streamlines shown), it creates a hot shock front onto the surface of the circumplanetary disk (dark orange surface above and below the planet). In this figure the circumplanetary disk is clearly visible with orange colors, showing that the disk is hotter than its surroundings.}
\label{fig:tempbig}
\end{figure*}

\begin{figure*}
\includegraphics[width=0.33\textwidth]{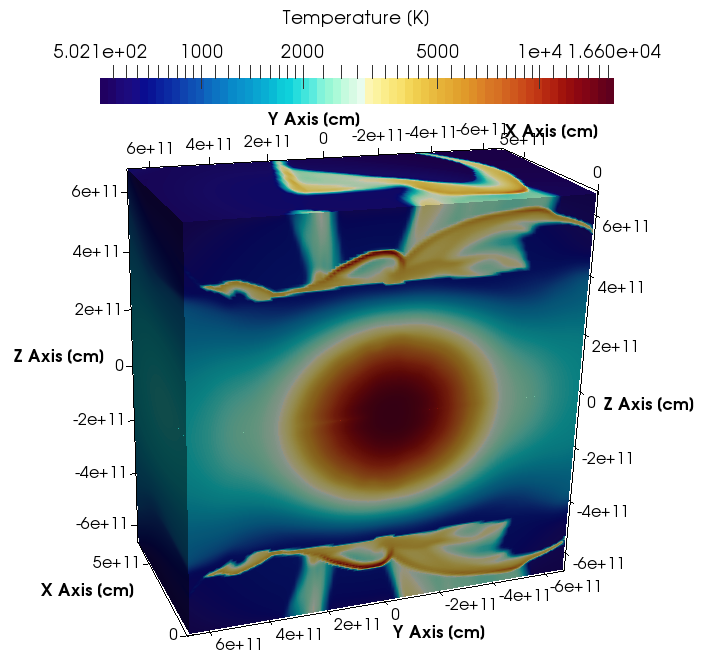}
\includegraphics[width=0.33\textwidth]{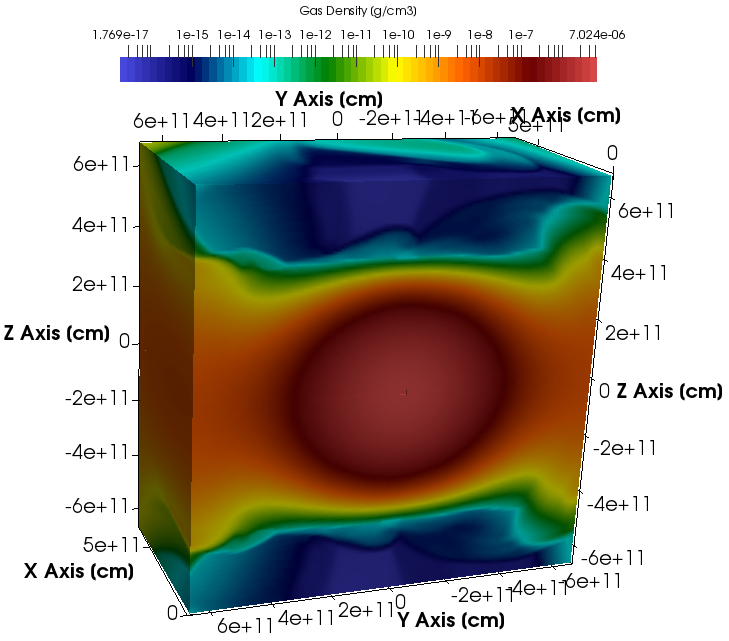}
\includegraphics[width=0.3\textwidth]{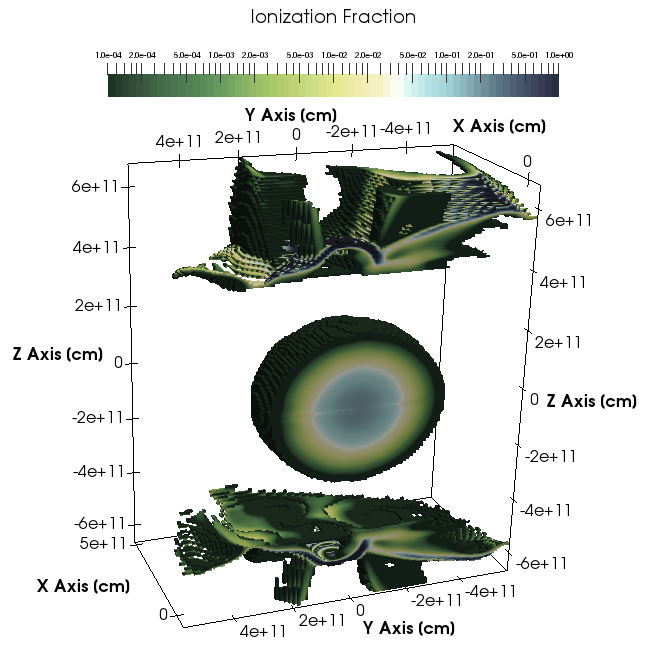}
\caption{Left: Temperature-map zoom into the planet vicinity for the 10 $\rm{M_{Jup}}$ simulation (i.e. the hottest case). The shock front on the circumplanetary disk above the planet is clearly visible with high temperature. This is the region which is hot enough to ionize hydrogen. Middle: Density color map of the same region. Right: ionization fractions in the same region. Clearly, ionization, and therefore hydrogen recombination line production happens in the planet and in the circumplanetary shock front. The planet's contribution to the line luminosity is, however, obscured  by the upper layers of the circumplanetary disk, meaning that only the line emission from the circumplanetary disk can be observable.}
\label{fig:ion}
\end{figure*}

One important result from 3D simulations is that the accretion shock-front is located on the circumplanetary disk \citep{SzM17,Tanigawa12}, rather than being on the planetary surface, as  suggested by one-dimensional simulations \citep{Marleau19}. Indeed 3D simulations show that the accretion stream is launched from the upper layers of the circumstellar disk, spiraling down to the circumplanetary disk through the so-called meridional circulation (Fig. \ref{fig:merid}; \citealt{Szulagyi14,FC16}, that has now been observed: \citealt{Teague19}) and hitting its surface, while creating a luminous shock-front on it (Figures \ref{fig:merid}, \ref{fig:streams}). This shock can be hot enough to ionize hydrogen (Figs. \ref{fig:tempbig}, \ref{fig:ion}), thus H-$\alpha$ emission could be expected to arise from both the planet and the circumplanetary disk. \citet{Aoyama18} carried out detailed one-dimensional analytical calculations in order to estimate the hydrogen recombination line fluxes under the assumption that the lines are emitted from the surface of forming planets. These calculations found that forming planets can indeed produce all the hydrogen recombination lines with very high line luminosities, again in tension with current observational results. This result was mainly driven by an assumption of extremely high temperatures (approx. $10^4-10^6$ Kelvin on the surface of planets) and high collisional velocities ($>20$ km/s), which are not expected on the surfaces of forming planets \citep{Szulagyi17}, but could be perhaps more appropriate for stars.

In this work we have self-consistently calculated hydrogen recombination line emissions from 3D thermo-hydrodynamical models of forming planets, including extinction. We examined the expected line variability for H-$\alpha$, Paschen-$\beta$ and Brackett-$\gamma$ lines and for the first time determined the parametric equation between the accretion luminosity $L_{acc}$ and the hydrogen recombination line fluxes for planets.

\section[]{Methods}
\label{sec:numerical}
\subsection[]{Hydrodynamical Simulations}
\label{sec:hydro}

We have used 3D thermo-hydrodynamical simulations performed with the JUPITER code \citep{Szulagyi16}. This algorithm, developed by F. Masset and J. Szulagyi solves the Euler equations, and calculates the temperature via a radiative module using a flux limited approximation (see e.g. on Fig. \ref{fig:tempbig}; \citealt{Kley89,Commercon11}). The heating and cooling channels include viscous heating, shock heating, adiabatic compression (e.g. due to accretion onto the planet), adiabatic expansion and radiative dissipation.

\begin{figure*}
\includegraphics[width=\textwidth]{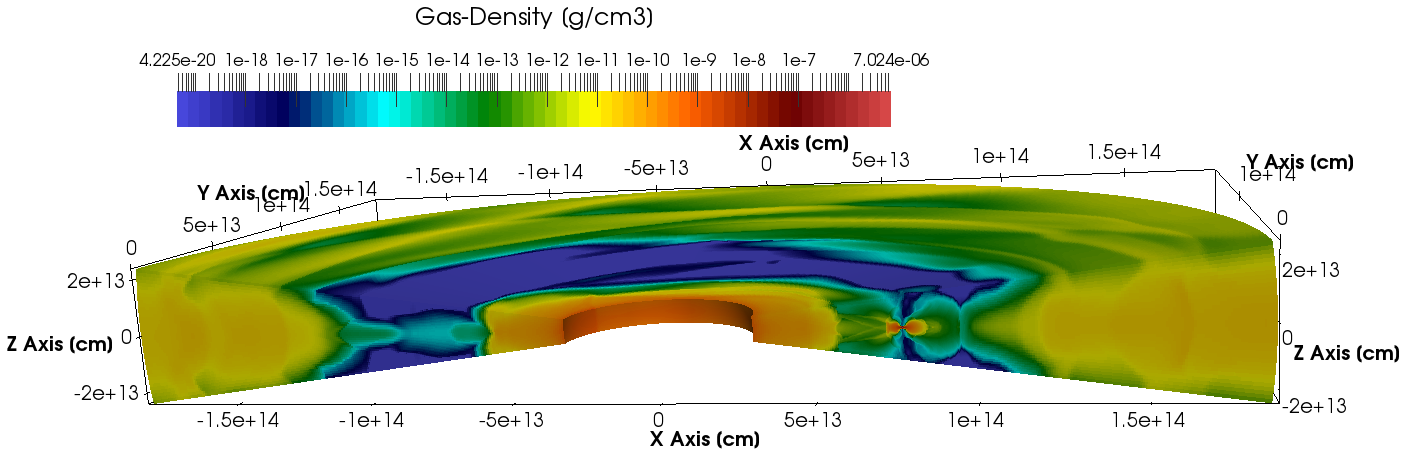}
\caption{The entire simulation box with the circumstellar disk for the 10 $\rm{M_{Jup}}$ planet case. There is an orange butterfly structure within the planetary gap in the right side of the disk: this is the circumplanetary disk developed around the forming planet.}
\label{fig:csd}
\end{figure*}

The simulations are the same as in \citet{SzM17} and consist of a circumstellar disk (Fig. \ref{fig:csd}) forming one planet of a given mass. The coordinate system is spherical and centered on the 1 Solar-mass star, that was treated as a point-mass. The ring of the circumstellar disk around the star spans a distance between 2.0 and 12.4 AU (sampled over 215 cells), the planet is placed at 5.2 AU (Jupiter's distance from the Sun). The circumstellar disk’s initial surface density is set as $\Sigma=\Sigma_0 (r/5.2AU)^{-0.5}$ with $\Sigma_0=222.2$ $\rm{g/cm^2}$. This way the total disk mass is approximately 11 $\rm{M_{Jup}}$, which is an average protoplanetary disk mass from observations \citep{WC11}. A power-law slope of -0.5 is again chosen based on observational constraints \citep{Andrews09,Isella09}. In different simulations we study planets of different masses, namely 1, 3, 5, and 10 $\rm{M_{Jup}}$. These are treated as a point-mass in the corner of 8 cells. The planet is thus unresolved and represented by a gravitational potential well. However, a small gravitational softening\footnote {gravitational softening means that within the smoothing radius ($r_s$) the planet's gravitational potential ($U_p$) is artifically lowered. The used equation is: $U_p=-\frac{G M_p}{\sqrt{x_d^2+y_d^2+z_d^2+{r_s}^2}}$} is used in order to avoid a singularity, with smoothing lengths of $6.5\times10^{10}$, $1.3\times10^{11}$, $1.3\times10^{11}$, and $2.7\times10^{11}$ cm for the 1, 3, 5, and 10 $\rm{M_{Jup}}$ planets, respectively. Nested meshes are used in the JUPITER code in order to increase the resolution near the planet. With each level of refinement, the resolution doubles and we use 6 levels of refinement to approach a length compared to Jupiter's radius, with the smallest cells being $1.1\times10^{10}$ cm, i.e. $\sim$ 0.8 Jupiter-diameter. The opening angle of the circumstellar disk is 7.4 degrees (using 20 cells resolution on the base mesh). The circumstellar disk is resolved with 680 cell to cover the $2\pi$ azimuthal extension. 
We employ the ideal gas equation of state with a fixed adiabatic index of 1.4 and a mean molecular weight of 2.3 (corresponding to the solar value). The fixed adiabatic index is a limitation, as in reality it would change due to molecular hydrogen dissociation and ionization. A constant kinematic viscosity of value of $10^{15}$ $\rm{cm^2/s}$ is used. The opacity table used in the hydrodynamic simulations is a frequency independent Rosseland-mean-opacity \citep{BL94}.

\subsection[]{Line luminosity calculation}
\label{sec:MOCASSIN}

We post-processed the temperature and density fields from the hydrodynamic simulations in order to calculate an ionization rate via the Saha-equation \citep{Saha20} and used this to derive the electron number densities. We obtained the hydrogen number densities from the total gas densities assuming a mean-molecular weight of 2.3, consistent with the hydrodynamic simulations. The hydrogen recombination spectrum was then calculated by applying the formalism used in the MOCASSIN code \citep{Ercolano03,Ercolano05,Ercolano08} and based on the detailed atomic data calculations of (Storey and Hummer 1995). We obtained the emerging line intensities of H-$\alpha$ (0.6563 $\mu$m), Paschen-$\beta$ (1.2818 $\mu$m) and Brackett-$\gamma$ (2.165 $\mu$m) by integrating the local line luminosities from the midplane of the disk to the surface, assuming the disk to be face-on (best case scenario) and including extinction along the line of sight, i.e. absorption due to the gas and the dust grains in the disk. The size distribution and the chemical composition of the dust grains in the disks are unfortunately poorly known and these can strongly affect the opacities. In order to assess the uncertainties introduced by the unknown grain properties we experimented with four very different opacity tables: 
\begin{enumerate}
\item dust mixture of 40\%  silicates + 40\% water-ice + 20\% carbon (i.e. graphite), with dust grain sizes of 1 microns, and dust-to-gas ratio of 1\% \citep{Draine03,Zubko96,WB08}.
\item silicate grains, with size distribution between 0.005 microns and 2.5 microns (with a slope of -3.5), and dust-to-gas ratio of 1\% \citep{DL84,Mathis77}).
\item graphite grains, with dust grain size distribution between 0.005 microns and 2.5 microns (with a slope of -3.5), and dust-to-gas ratio of 1\% \citep{Draine03,Mathis77}.
\item only gas opacities \citep{Draine03}, describing a case where the dust has sublimated or the disk is otherwise depleted from dust.
\end{enumerate}
The gas-only case gives an absolute upper limit (theoretical limit) for the line luminosities, since the presence of dust in protoplanetary disks is well-established from observations. The graphite (i.e. carbon) grains have very high opacity, and considering only those is just a limit of highest possible opacity-case (i.e. minimal line luminosities). A realistic disk might have a composition similar to our silicate-water-carbon case. 

Finally, we estimated the variability of the lines. The main reasons for variability are the changes in extinction and changes in the accretion rate. In the hydrodynamic simulation there are changes in the CPD accretion rate from the meridional circulation as the planet orbits around the star. This leads to changes in the strength/temperature of the shock front on the circumplanetary disk (Fig. \ref{fig:tempbig}). Similarly, due to the orbit around the star, and the CPD rotation around the planet, the extinction column changes rapidly (see more in Sect. \ref{sec:variab}).

\section{Results}

\subsection{The Line Luminosities}
\label{sec:line}
The computed line luminosities in Solar-luminosity units ($\rm{L_{\odot}}$) can be found in Table \ref{tab:linelum}. We tested what percentage of the line luminosities comes from the circumplanetary disk only, by removing the planet entirely from the integration for the line emissivities. This calculation showed that 100\% of the reported luminosities come from the circumplanetary disk itself in our boundary-layer accretion assumption. If the planet has strong magnetic fields, this could lead to a cavity between the planet surface and the CPD, that would help the planet-generated hydrogen recombination lines to escape. Therefore, in the boundary-layer accretion assumption, the observability of hydrogen lines from forming planets can only be estimated via modeling of the circumplanetary disk rather than the planet interior and surface. 

With the most realistic disk opacities (the silicate-water-carbon mixture), only the 10 $\rm{M_{Jup}}$ case could be robustly observed with luminosities peaking at $\sim$few times $10^{-2}\, \rm{L_{\odot}}$ for H-$\alpha$ and for Paschen-$\beta$ and $\sim$few times $10^{-3}\, \rm{L_{\odot}}$ for the Brackett-$\gamma$ lines. For the 5 $\rm{M_{Jup}}$ due to the line variability, sometimes the hydrogen recombination line luminosities are on the detection limit with values between $\sim 10^{-8} \rm{ and } 10^{-5}\, \rm{L_{\odot}}$. We emphasize, however, these calculations assume the most favorable configuration (face-on disk, local-thermal equilibrium approximation), and should thus be considered as upper limits. For the lower mass planet cases ($\leq$ 3 $\rm{M_{Jup}}$ ) with this opacity table, the line luminosities are always $<10^{-11} \, \rm{L_{\odot}}$ for all lines, well below the detection limits of current instruments. This might explain why only so few detections of H-$\alpha$ from planetary mass objects exist, given that planets above 5 $\rm{M_{Jup}}$ are rare.

For comparison purposes, we have also calculated the line luminosities with pure silicate opacities only. As silicate has lower opacity than carbon and water, these line fluxes are higher than the mixture case containing those two (Table \ref{tab:linelum}). For the 3 $\rm{M_{Jup}}$ and 5 $\rm{M_{Jup}}$ cases the line luminosities usually range between $10^{-4} \, \rm{L_{\odot}}$ to $10^{-9} \, \rm{L_{\odot}}$, which could be observable with current instruments in favorable conditions. The 10 $
\rm{M_{Jup}}$ planet produces between $10^{-3} \rm{ and } 10^{-2}\, \rm{L_{\odot}}$ for Paschen-$\beta$ and Brackett-$\gamma$ lines and  up to $10^{-1} \, \rm{L_{\odot}}$ for H-$\alpha$. Using the graphite (i.e. carbon) opacity, the line luminosities are more extinct than in the pure silicate case. Still, the 5 Jupiter-mass cases could be near the observable limits for all lines (Table \ref{tab:linelum}). All the hydrogen recombination lines are easily observable from the 10 Jupiter-mass planet simulations, however, the 1-3 $\rm{M_{Jup}}$ cases are below the detection limit.

Considering only gas opacities (i.e. almost no extinction), all the lines for all planetary masses considered here could be observable with line fluxes larger than $10^{-6} \, \rm{L_{\odot}}$. This case might be relevant only if all dust grains have sublimated or in the unlikely case of circumplanetary disks that are intrinsically free of dust \citep{Zhu15}. These line fluxes obtained under the no dust assumption, therefore should be considered as theoretical upper limits. More likely dust extinction in the circumplanetary disk \citep{DSz18} and its atmosphere will reduce the line luminosities by several orders of magnitude (see Table  \ref{tab:linelum}. 

Our results suggest that when realistic opacities are considered (mixture of silicate, water and carbon), detecting forming planets smaller than 10 $\rm{M_{Jup}}$ via hydrogen recombination lines is challenging with current instrumentation. Forming planets larger than this limit might be observable, but unfortunately their occurrence rate is relatively low\footnote{\url{http://exoplanet.eu/}}. As well as intrinsically brighter line luminosities, the observability of 10 $\rm{M_{Jup}}$ forming planets is higher due the planet carving a deeper gap in the circumstellar disk which results in lower extinction above the circumplanetary disk shock-surface. In the lower planetary mass cases, the gap is shallower and less empty, hence extinction more severely affects the line luminosities. The situation is of course more complex when the system is inclined from the line of sight, since in this case additional line absorption from circumstellar disk material can then be expected as well.

\begin{deluxetable*}{l|c|cccc|cccc|cccc}
\rotate
\tablecolumns{14}
\tabletypesize{\scriptsize}
\tablehead{
\colhead{}& \colhead{1 $\rm{M_{Jup}}$} & \multicolumn{4}{c}{3 $\rm{M_{Jup}}$} & \multicolumn{4}{c}{5 $\rm{M_{Jup}}$} & \multicolumn{4}{c}{10 $\rm{M_{Jup}}$}}
\startdata
$\rm{L_{H\alpha}}$ (mix) & 0.00E+00 &  1.09E-24 &  1.82E-22 &  0.00E+00 &  6.11E-11 &  3.93E-16 &  6.42E-06 &  3.42E-05 &  4.32E-11 &  6.32E-02 &  3.40E-02 &  1.61E-01 &  6.20E-02   \\
$\rm{L_{Br\gamma}}$ (mix) & 0.00E+00 &  2.93E-27 &  5.97E-25 &  0.00E+00 &  5.54E-13 &  2.60E-18 &  9.46E-08 &  4.90E-07 &  4.42E-13 &  9.46E-04 &  4.22E-04 &  2.20E-03 &  9.72E-04   \\
$\rm{L_{Pa\beta}}$ (mix) & 0.00E+00 &  1.61E-25 &  2.40E-23 &  0.00E+00 &  4.90E-12 &  4.65E-17 &  4.91E-07 &  2.98E-06 &  3.06E-12 &  5.40E-03 &  2.89E-03 &  1.39E-02 &  4.91E-03  \\
$\rm{L_{H\alpha}}$ (silicate) & 0.00E+00 &  2.62E-09 &  8.28E-09 &  3.88E-15 &  1.27E-06 &  1.19E-06 &  2.63E-04 &  3.99E-03 &  1.54E-07 &  1.07E-01 &  6.55E-02 &  1.83E-01 &  6.68E-02   \\
$\rm{L_{Br\gamma}}$ (silicate) & 2.32E-11 &  1.13E-06 &  2.05E-06 &  1.56E-06 &  1.16E-06 &  1.26E-04 &  7.77E-05 &  3.63E-04 &  7.40E-05 &  2.17E-03 &  1.13E-03 &  2.61E-03 &  1.10E-03   \\
$\rm{L_{Pa\beta}}$ (silicate) & 2.21E-23 &  7.60E-07 &  1.60E-06 &  1.38E-07 &  2.12E-06 &  9.31E-05 &  1.31E-04 &  1.20E-03 &  1.16E-05 &  1.08E-02 &  6.99E-03 &  1.62E-02 &  5.43E-03   \\
$\rm{L_{H\alpha}}$ (graphite) & 0.00E+00 &  8.94E-20 &  3.75E-18 &  0.00E+00 &  1.30E-09 &  2.85E-13 &  1.75E-05 &  1.42E-04 &  3.13E-10 &  7.21E-02 &  4.05E-02 &  1.68E-01 &  6.32E-02  \\
$\rm{L_{Br\gamma}}$ (graphite) & 0.00E+00 &  7.47E-10 &  1.92E-09 &  7.97E-14 &  4.93E-08 &  1.85E-07 &  7.33E-06 &  1.13E-04 &  6.20E-09 &  1.82E-03 &  9.12E-04 &  2.55E-03 &  1.06E-03   \\
$\rm{L_{Pa\beta}}$ (graphite) & 0.00E+00 &  2.37E-13 &  1.38E-12 &  1.97E-23 &  1.12E-08 &  9.96E-10 &  7.84E-06 &  1.18E-04 &  1.42E-09 &  7.89E-03 &  4.75E-03 &  1.54E-02 &  5.18E-03  \\
$\rm{L_{H\alpha}}$ (gas)&  7.95E-05 &  1.89E-04 &  2.62E-04 &  4.85E-04 &  1.11E-04 &  1.64E-02 &  7.70E-03 &  2.29E-02 &  1.93E-02 &  1.36E-01 &  8.78E-02 &  1.89E-01 &  6.95E-02   \\
$\rm{L_{Br\gamma}}$ (gas) & 1.64E-06 &  3.08E-06 &  4.21E-06 &  8.13E-06 &  1.71E-06 &  2.95E-04 &  1.37E-04 &  4.09E-04 &  3.48E-04 &  2.20E-03 &  1.15E-03 &  2.61E-03 &  1.10E-03   \\
$\rm{L_{Pa\beta}}$ (gas) & 6.24E-06 &  1.29E-05 &  1.78E-05 &  3.37E-05 &  7.38E-06 &  1.27E-03 &  5.71E-04 &  1.78E-03 &  1.49E-03 &  1.13E-02 &  7.46E-03 &  1.63E-02 &  5.48E-03   \\
Mass influx rate  & 2.61E-05 &  4.76E-05 &  3.66E-05 &  4.25E-05 &  2.66E-05 &  8.44E-05 &  6.73E-05 &  7.16E-05 &  9.22E-05 &  5.51E-05 &  4.80E-05 &  4.58E-05 &  1.18E-04\\
to CPD $[\rm{M_{Jup}/year}]$ &&&&&&&&&&&&&\\
\hline
\enddata
\caption{Line luminosities in Solar-luminosity [$L_{\odot}$] units for each planetary masses, and four different opacity tables (indicated in the brackets in the first column). The four values given for the 3-10 $
\rm{M_{Jup}}$ measured at four points of the orbit of the planet around the star, to examine line variability. The last row is the mass influx rate to the CPD, highlighting the chances in this value, that also contribute to the line variability, apart from changes in the extinction column.}
\label{tab:linelum}
\end{deluxetable*}

\subsection{Variability}
\label{sec:variab}

The line variability was also examined. As mentioned before, the accretion rate onto the CPD and thus the shock-front strength will change as the planet orbits around the star. Even more importantly, the density around the planet \& CPD changes as well during the orbit, leading to different extinction columns at different times. For these reasons, we considered four different points during the orbit of the planet, whenever such simulation outputs were available (for the 3, 5, 10 Jupiter-mass simulations). We calculated the line variability using the data from Table \ref{tab:linelum}, excluding the non-detections (defined by $<10^{-27}\, \rm{L_{\odot}}$). The standard-deviation of the four values in each cases were divided with the maximal line-luminosity, and expressed in percentages (Table \ref{tab:vari}). We used the maximal line-luminosity in each case instead of the average, because due to the often low luminosity values, the maximums are more robust than the averages.

In most cases, the variability is between 28-58\,\% (Table \ref{tab:vari}). These are slightly higher values than the corresponding variability of the mass influx to the CPD. The variability of the influx is 18.9\,\%, 12.4\,\%,   19.1\,\%  for the 3, 5, 10 Jupiter-mass cases, respectively. This highlights that the reason for the variability are partially the mass influx rate to the CPD (and hence the varied shock-front strengths) and the changes in the extinction column. Furthermore, the effect of variable extinction can be seen by the lower line variability values in the gas-only cases in Tab. \ref{tab:vari} than in the dust-included opacity cases. The high variability of these lines can be a further explanation why in the case of the LkCa 15b planet candidate, H-$\alpha$ could be detected in one observation \citep{Sallum15}, but not in others \citep{Mendigutia18}; it is indeed possible that the H-$\alpha$ production at the time of the second observation was just below the observational limit.

 \begin{table}
  \begin{tabular}{l|c|c|c}
\hline
    &	3 $
\rm{M_{Jup}}$ [\%] &	5 $
\rm{M_{Jup}}$ [\%]	 &10 $
\rm{M_{Jup}}$ [\%]\\
     \hline
$\rm{\Delta_{H\alpha}}$ (mix) &57.7 & 47.7 & 34.6   \\
$\rm{\Delta_{Br\gamma}}$ (mix) &57.7 & 47.7 & 34.3   \\
$\rm{\Delta_{Pa\beta}}$ (mix) &57.7 & 47.9 & 35.1   \\
$\rm{\Delta_{H\alpha}}$ (silicate) &49.9 & 49.0 & 30.1   \\
$\rm{\Delta_{Br\gamma}}$ (silicate) &20.9 & 37.8 & 29.0   \\
$\rm{\Delta_{Pa\beta}}$ (silicate) &41.5 & 46.9 & 29.6  \\
$\rm{\Delta_{H\alpha}}$ (graphite) &57.7 & 48.3 & 33.5  \\
$\rm{\Delta_{Br\gamma}}$ (graphite) &49.1 & 49.0 & 29.6   \\
$\rm{\Delta_{Pa\beta}}$ (graphite) &50.0 & 49.0 & 32.0   \\
$\rm{\Delta_{H\alpha}}$ (gas) &33.2 & 28.3 & 28.4   \\
$\rm{\Delta_{Br\gamma}}$ (gas) &34.0 & 28.5 & 29.0   \\
$\rm{\Delta_{Pa\beta}}$ (gas) &33.7 & 29.0 & 29.3   \\
 \hline
\end{tabular}
  \caption{Variability of the lines (in percentages) for the four different opacity tables and the various planetary masses.}
   \label{tab:vari}
\end{table}

\subsection{The relation between ${M_{planet}}$ and $L_{Hline}$}
\label{sec:formulaMplanet}

In general, the line luminosities increase with increasing planetary mass (Tab. \ref{tab:linelum}). We calculated the regression between each line versus the planetary mass, to obtain a rough $\rm{M_{planet}} \propto log_{10}(\rm{L_{line}})$ relationship. We used the data from Table \ref{tab:linelum}, excluding the non-detection cases when calculating the regression. The coefficients of the fitted lines can be found in Table \ref{tab:pmass} for each opacity case separately. The trend of increasing line luminosity with increasing planetary mass is on one hand driven by the temperature being generally higher in the vicinity of a more massive planet, resulting in more hydrogen ionization and hence higher line luminosities. Furthermore, the extinction column decreases as well with increasing planetary mass: the larger mass planets open deeper gaps, and the atmosphere region of the CPD is thinner (while the CPD density is overall higher), than in the low-planet mass cases. We also found that all of the observable hydrogen recombination line production comes from the circumplanetary disk shock, rather than from the planet itself. An observer might see this luminous shock front, while the planet emissivity would be absorbed by the large extinction in the circumplanetary disk (Fig. \ref{fig:ion}). 

 \begin{table}
  \begin{tabular}{lccc}
  \hline
$\rm{L_{line}}$ (opacity table)&	a	& b &	$\sigma$\\
  \hline
$\rm{L_{H\alpha}}$ (mix)	& 	2.23  &  -22.76  &  0.53 \\
$\rm{L_{Br\gamma}}$ (mix)	& 	2.29  &  -25.16  &  0.55 \\
$\rm{L_{Pa\beta}}$ (mix)	& 	 2.22  &  -23.69  &  0.52 \\
$\rm{L_{H\alpha}}$ (sil.)	&  1.09 &  -11.51 &  0.24 \\
$\rm{L_{Br\gamma}}$ (sil.)	& 	0.55  &  -7.80  &  0.11 \\
$\rm{L_{Pa\beta}}$ (sil.)	& 	1.12  &  -11.75  &  0.37 \\
	$\rm{L_{H\alpha}}$ (graph.)	& 	1.81  &  -18.68  &  0.40 \\
$\rm{L_{Br\gamma}}$ (graph.)	& 	0.89  &  -11.52  &  0.18 \\
	$\rm{L_{Pa\beta}}$ (graph.)	&  1.52  &  -16.67  &  0.40 \\
$\rm{L_{H\alpha}}$ (gas)	& 	 0.35  &  -4.26   &    0.04 \\
$\rm{L_{Br\gamma}}$ (gas)	& 0.34  &  -5.99  &    0.05 \\
$\rm{L_{Pa\beta}}$ (gas)	& 	0.36  &  -5.43  &    0.04 \\
 \hline
\end{tabular}
  \caption{Planet mass vs. hydrogen recombination line luminosities. Coefficients for regression for $\log_{10}(L_{line})=a*M_{planet}+b$ relation and the 1-sigma uncertainty estimates for the parameter "$a$". The four different opacity tables considered are indicated in the parenthesis in column one. }
   \label{tab:pmass}
\end{table}

\subsection{The relation between $L_{acc}$ and $L_{Hline}$}
\label{sec:formula}

As mentioned above, an empirical relation between accretion luminosity ($\rm{L_{acc}}$) and the hydrogen recombination line luminosity ($\rm{L_{Hline}}$) has been derived from observations of young T Tauri stars 
\citep{Rigliaco,Alcala14}. However the validity of this relationship for planets is questionable. In what follows we determine the relationship between $\rm{L_{acc}}$ and $\rm{L_{Hline}}$ based on our simulations. First, we obtained the mass influx rate to the circumplanetary disk (\citealt{Tanigawa12,Szulagyi14} $A=\rho s v_z$, where $\rho$ stands for density, $s$ for surface, and $v_z$ for z-component of velocity) that generates the shock front (see last row in Table \ref{tab:linelum}). This is not the actual accretion (i.e. net growing) rate of the circumplanetary disk, or of the planet, because most of this mass ($>90$\%) will be recycled and will flow back to the midplane regions of the circumplanetary disk (Fig. \ref{fig:streams}, \ref{fig:merid}). The accretion luminosity can be computed from the mass fluxes using the formula: $L_{acc} = GM_{planet}A/R$, where $G$ is the gravitational constant, $A$ is the mass influx rate and $R$ is the distance of the surface where the mass influx was computed. We used $R = 4\times10^{11}$ cm, which is just above the shock-front in these simulations. As a next step, we computed the regression between $\log_{10}(L_{acc})$ and  $\log_{10}(L_{Hline})$ from Table \ref{tab:linelum}, excluding the non-detection cases. The fitted line values and uncertainties are given in Table \ref{tab:lacc}. Here, the slope seems to be steeper for planets than in the case of stars (slope $\cong 1$ \citealt{Rigliaco,Alcala14}), suggesting  that the accretion process might indeed be different for planets and for stars. However, this is not surprising since we have assumed  boundary-layer accretion in our calculations. It remains to be seen whether calculations including magnetospheric accretion might retrieve slopes closer to the T Tauri values. In any case, our values also indicate that, in the case of boundary-layer accretion, previous observations and analyses which employed the T Tauri formula for forming planets might have overestimated the accretion rates of the planets (or rather, of the circumplanetary disks). 	

 \begin{table}
  \begin{tabular}{lccc}
  \hline
$\rm{L_{line}}$ (opacity table)&	a	& b & 	$\sigma$\\
  \hline
$\rm{L_{H\alpha}}$(mix) & 	17.06 &  38.90 &  5.90 \\
$\rm{L_{Br\gamma}}$(mix) & 	 17.59 &  38.37 &  6.08 \\
$\rm{L_{Pa\beta}}$(mix) & 	16.91 &  37.46 &  5.83 \\
$\rm{L_{H\alpha}}$(sil.) & 	 8.45 &  18.91 &  2.65 \\
$\rm{L_{Br\gamma}}$(sil.) & 	 4.63 &  8.79 & 0.53 \\
$\rm{L_{Pa\beta}}$(sil.) & 	 10.14 &  24.03 &  2.08 \\
$\rm{L_{H\alpha}}$(graph.) & 	13.60 &  30.65 &  4.65 \\
$\rm{L_{Br\gamma}}$(graph.) & 	6.90 &  13.37 &  2.06 \\
$\rm{L_{Pa\beta}}$(graph.) & 	12.07 &  26.58 &  4.08 \\
$\rm{L_{H\alpha}}$(gas) & 	 2.58 &  5.23 & 0.35 \\
$\rm{L_{Br\gamma}}$(gas) & 	2.53 &  3.31 & 0.35 \\
$\rm{L_{Pa\beta}}$(gas) & 	2.63 &  4.25 & 0.37 \\
 \hline
\end{tabular}
  \caption{Accretion luminosity vs. hydrogen recombination line luminosities. Coefficients for regression for  $\log_{10}(L_{line})=a*\log_{10}(L_{acc})+b$ relation and the  1-sigma uncertainty estimates for the parameter "$a$". The four different opacity tables are marked in the parenthesis after each line luminosity.}
   \label{tab:lacc}
\end{table}

\section{Discussion}

These simulations do not contain magnetic fields, however the circumplanetary disk magnetic field could change the accretion rate to the planet \citep{Gressel13}. If the planet  has strong magnetic fields ($\geq$ 65 Gauss - according to \citealt{OM16}), it can accrete via magnetospheric accretion like stars do, instead of boundary layer accretion \citep{Batygin18}. Due to lack of simulations on the topic, it is unknown what field strength forming planets might have and whether they accrete via boundary layer accretion or magnetospheric accretion. It could also very well be, that this is not a "this or that" question, and some planets would accrete one way, while others the other way. Furthermore, the magnetic field of the planet is expected to be changing during its evolution, like it is the case for stars \citep{Christensen}. Clearly, magnetospheric accretion would completely change the accretion flow and -rate in comparison to what this work simulations show. In any case, for magnetospheric accretion to happen, two conditions need to be met: first, the planet should have strong ($\geq$ 65 Gauss - \citealt{OM16}) field; second, there should be sufficient amount of ionized gas in the CPD. Regarding the first, Jupiter has today 5 Gauss field. It could be that planets during their formation had larger magnetic field, as this is true for stars \citep{Christensen}. \citet{OM16} back-of-an-envelope estimation suggested that Jupiter might have had $\sim 50$ Gauss field during its infancy, but this would still not be enough to launch magnetospheric accretion. Regarding the ionization rates, the bulk of the CPD is neutral (see e.g. Fig. \ref{fig:ion}; \citealt{Fujii11,Fujii14}). However, \citet{Batygin18} calculated from our models that alkali elements might have enough ionization. In any case, full 3D magneto-hydrodynamic simulations are needed with planet magnetic fields included, in order to understand how the accretion stream changes in the magnetospheric accretion case. So far no such simulation was possible due to the limitations of the today's computers. Furthermore, dynamo models are needed for forming planets to understand what field strength we can expect in this stage (that is still completely unknown). It is likely, that our parametric equation slope value differs from that of stars, because of our assumption of boundary layer accretion. The parametric equations would be likely different for magnetospherically accreting planets.

In this work we also omitted gas self-gravity, that could play an important role in the runaway phase \citep{Pollack96,Piso15}. The inclusion of self-gravity would help on the CPD-collapse and presumably increase the accretion rate \citep{AB12}. However running 3D global simulations with radiation transport, as well as with planet-resolution in the Hill-sphere, together with self-gravity, would be computationally very expensive and could be done only on the next generation computers.

The accretion rates of planets and CPDs are known to depend on the circumstellar disk surface density, the viscosity assumed, the scale-height of the circumstellar disk (e.g. the semi-major axis of the planet, the temperature of the CSD). These effects therefore can also affect somewhat the calculated line fluxes, (in particular, the different density will lead to different extinction rate. Due to the expensiveness of these high resolution simulations with all the physics mentioned in Sect. \ref{sec:hydro} included (approx. $1.2\times 10^4$ core-hours per simulation), it is only possible to explore a small part of the parameter space.

Our simulations had a resolution somewhat smaller than the radius of forming planets ($\sim$0.8 Jupiter-diameter), and we did not have any planet interior model included. Such models could potentially better estimate the hydrogen recombination line emission from the forming planets. The emission of the interior, however, necessarily will be absorbed by the interior and outer regions of the planet. Only the atmosphere production would be able to leave the planet, but then -- as it happens in our simulations as well --, it will likely be absorbed by the CPD material. 

\section{Conclusion}

In this work we computed self-consistently H-$\alpha$, Paschen-$\beta$ and Brackett-$\gamma$ line luminosities from 3D thermo-hydrodynamical simulations with JUPITER \citep{Szulagyi16}. The planet masses considered were 1, 3, 5 and 10 Jupiter-masses. Line luminosities from recombination cascades were estimated using the formalism in the MOCASSIN code \citep{Ercolano03,Ercolano05,Ercolano08} in combination with a ionisation calculation using the Saha equation. We explored various opacities to estimate the line emissions with extinction in each cases. In the most realistic opacity case (dust mixture of 40\%  silicates + 40\% water-ice + 20\% carbon) only planets $\ge$10 Jupiter-mass can be detected with current instrumentation. The detectable line flux originate in all cases from the CPD surface shock-front, while the planet emitted contributions are completely absorbed. Our results indicate that detecting hydrogen recombination lines from forming planet is very challenging with current instrumentation, which might explain the very few detections reported so far. 

Our study on line flux variability showed a change of 28-58\,\%, mainly due to the changes in extinction (changes in density) as the planet orbits around the star and the CPD rotates around the planet. Secondly, the changes in the mass influx to the CPD along the orbit also changes the strength of the accretion shock, that translates into different line luminosities.

We determined for the first time the parametric equation between the accretion luminosity and the H recombination line luminosity for planets. Moreover, we also determined the relationship between the planetary mass and the line luminosities for H-$\alpha$, Paschen-$\beta$ and Brackett-$\gamma$ separately, and for the various opacity cases. These relationships for planets seem to be steeper than the equations for T Tauri stars, with our boundary-layer accretion assumption. Future magneto-hydrodynamic simulations are needed in the CPD region to examine a potential magnetospheric accretion if the planet magnetic fields are very high $>65\,$Gauss \citep{OM16}.

\acknowledgments
We thank to Mickael Bonnefoy for his comments with the observational aspects, and for the anonymous referee for the useful suggestions. This work has been in part carried out within the Swiss National Science Foundation (SNSF) Ambizione grant PZ00P2\_174115. B.E. acknowledges support from the DFG Research Unit “Transition Disks" (FOR 2634/1, ER 685/8-1). Computations have been done on the "Piz Daint" machine hosted at the Swiss National Computational Centre.

\bibliography{ref}

\end{document}